\documentclass[%
 reprint,
 amsmath,amssymb,
 aps,
]{revtex4-1}

\usepackage{graphicx}
\usepackage{dcolumn}
\usepackage{bm}

\usepackage[utf8]{inputenc}
\usepackage[T1]{fontenc}
\usepackage{here}

\begin{document}


\title{Characterization of weak deep UV pulses using cross-phase modulation scans}

\author{Jan Reisl\"ohner}
\author{Christoph Leithold}
\author{Adrian N. Pfeiffer}

\affiliation{Institute of Optics and Quantum Electronics, Abbe Center of Photonics, Friedrich Schiller University, Max-Wien-Platz 1, 07743 Jena, Germany}


\date{\today}

\begin{abstract}
Temporal pulse characterization methods can often not be applied to UV pulses due to the lack of suitable nonlinear crystals and very low pulse energies. Here, a method is introduced for the characterization of two unknown and independent laser pulses. The applicability is broad, but the method is especially useful for pulses in the deep UV, because pulse energies on the picojoule-scale suffice. The basis is a spectral analysis of the two interfering UV pulses, while one of the pulses is phase shifted by an unknown VIS-IR pulse via cross-phase modulation. The pulse retrieval is analytic and the fidelity can be checked by comparing the complex-valued data trace with the retrieved trace.
\end{abstract}

\maketitle



Few-cycle deep ultraviolet (DUV) pulses are used in various techniques of ultrafast spectroscopy \cite{Kobayashi1, Varillas, Stolow, Prokhorenko}. Temporal pulse characterization is often required, but experimental conditions impose restrictions on characterization methods. If a known reference pulse is available, linear characterization techniques of spectral interferometry can be used \cite{Lepetit}. Without a known reference, nonlinear characterization methods such as \textit{frequency resolved optical gating} (FROG) \cite{Kane} or \textit{spectral phase interferometry for direct electric-field reconstruction} (SPIDER) \cite{Iaconis} must be used. While these methods are very successful in the VIS-IR, their application in the DUV is often impossible because of low pulse energies and the lack of suitable nonlinear crystals. Intense pulses with microjoule-scale pulse energies can be characterized by frequency-degenerate methods such as self-diffraction FROG \cite{Durfee} or transient grating FROG \cite{Fuji,Nagy}. Also multiphoton processes in gas-phase media have been exploited for pulse characterization, for example in third-order fringe-resolved autocorrelation \cite{Reiter} or in cross-correlation \cite{Trushin}. This, however, requires intense pulses and the underlying nonlinear process is resonant, which complicates the pulse retrieval. For lower pulse energies, auxiliary VIS-IR pulses are often used to aid the nonlinear process, for example in zero-additional-phase SPIDER \cite{Baum} or in cross-correlation FROG (XFROG) \cite{Linden}. This, however, still requires nanojoule-scale pulse energies in the DUV \cite{Ermolov}.

Here, a method for the simultaneous characterization of two unknown and independent laser pulses is introduced and demonstrated in the DUV. Spectra of the two DUV pulses at a fixed pulse delay are recorded, while an unknown VIS-IR pulse is overlapped with one of the pulses at a variable delay. This shifts the phase of one DUV pulse via cross-phase modulation (XPM). Two variants of implementing the XPM scan are presented, named \emph{edge} and \emph{center}, which makes the method broadly applicable, including the case that all three pulses are frequency-degenerate. The recorded data trace contains both amplitude and phase information and allows an analytic pulse retrieval. At the same time, the fidelity of the retrieval can be checked by comparing the complex-valued data trace with the retrieved trace. This combines the advantage of rigour and speed of analytic retrieval, which is typical for SPIDER methods, with the reliability due to the feedback by a retrieved trace, which is typical for FROG methods. This combination of analytic retrieval and feedback is similar in \textit{interferometric imaging of self-diffraction}\,\cite{Leithold1, Leithold2}, a recently introduced method for the characterization of VIS-IR pulses. A further and very important advantage of the characterization using XPM scans is that pulse energies on the picojoule-scale suffice, which makes the method especially useful for pulses in the DUV. 

\begin{figure}[H] 
\centering
\includegraphics[width=\linewidth]{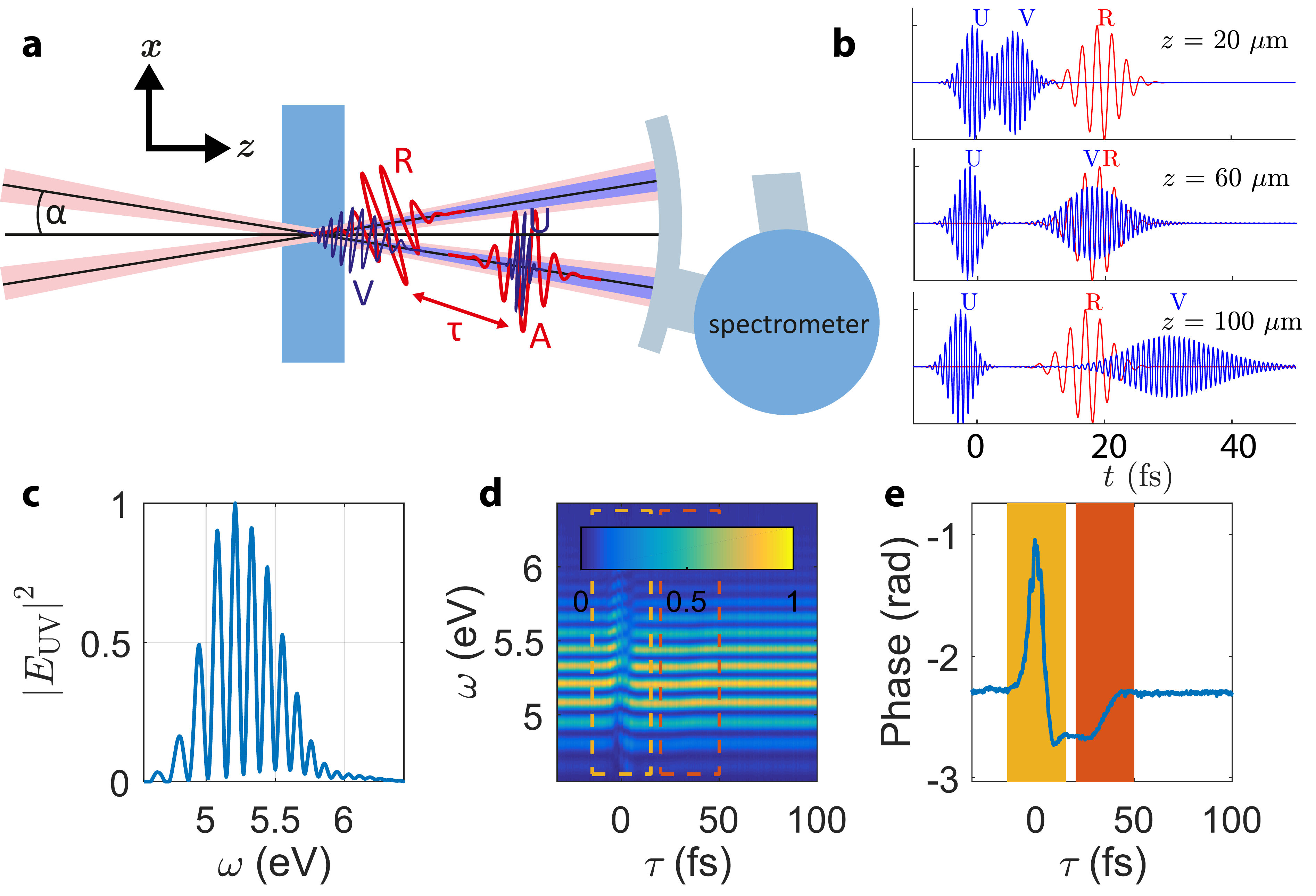}
\caption{Pulse A generates DUV pulses U and V while pulse R induces XPM at variable delay $\tau$ (a). The calculated and normalized electric fields during propagation are depicted in (b) for $\tau=20\ \mathrm{fs}$. The measured spectrum $ \left| E_{\mathrm{UV}}\right|^2 $ is shown in (c) for $\tau=100\ \mathrm{fs}$. The delay scan $\textbf{I}(\omega,\tau)$ is shown in (d). The phase of $\textbf{I}_r(\omega = 5.2\ \mathrm{eV},\tau)$ is plotted in (e). The delay range used in the variants \emph{edge} and \emph{center} are marked in red and yellow, respectively.}
\label{fig:1}
\end{figure}

Experimentally, two few-cycle VIS-IR pulses, labeled A and R, are focused noncollinearly (beam waist $\sim$\,100\,$\mu$m, half opening angle $\alpha = 0.55^{\circ}$) with polarization perpendicular to the plane of incidence into a nonlinear dielectric with variable delay $\tau$, see Fig.\,\ref{fig:1}(a). Here a 100-$\mu$m-thick amorphous MgF$_2$ plate is used, but any isotropic dielectric with a wide band gap is suited. A and B are similar as they originate by beam splitting of one laser pulse (center wavelength $\lambda_{A,R} = 700\ \mathrm{nm}$, pulse duration $t^{FWHM}_{A,R} = 6\ \mathrm{fs}$ and intensity $I_{A,R} = 4\ \mathrm{TW/cm}^2$), but this similarity is no prerequisite. DUV light is produced by third-order harmonic generation (THG) in MgF$_2$. The DUV light that is emitted collinearly to A is detected with a spectrometer.

The group velocities in the DUV and VIS-IR differ significantly (group index 1.39 @ $700\ \mathrm{nm}$, 1.50 @ $230\ \mathrm{nm}$). The 1D pulse propagation in $z$-direction is given by (atomic units are used and the convention for Fourier transform is $\mathcal{F} \{ f(t) \} \propto \int_{-\infty}^{+\infty} f(t) \mathrm{e}^{-i \omega t} \mathrm{d}t $):
\begin{equation} 
E(\omega,z) = \phi(\omega) \left( E_0(\omega) -  i\frac{2\pi\omega}{cn(\omega)}\int_{0}^{z} dz' P(\omega,z') \mathrm{e}^{ik(\omega)z'} \right),
\label{eq:PulseProp1D}
\end{equation}
where $\omega$ is the angular frequency, $n(\omega)$ is the refractive index and $k(\omega) = n(\omega) \frac{\omega}{c}$. $P$ is the nonlinear polarization response, $E_0(\omega) = E(\omega,z=0)$, and $\phi(\omega) = \mathrm{e}^{i\left(n_{IR} \frac{\omega}{c} - k(\omega)\right)z}$ describes the linear propagation in the reference frame of the IR phase velocity $c$/$n_{IR}$. With restriction to linear propagation at fundamental frequencies and with the assumption $n(\omega_{IR}) = n_{IR}$, Eq.\,(\ref{eq:PulseProp1D}) can be solved for DUV frequencies: $E_{\mathrm{UV}}(\omega,z) \propto (1 - \phi(\omega))$. For short fundamental laser pulses with a broad bandwidth, fringes appear in the spectrum $ \left| E_{\mathrm{UV}}\right|^2 $, see Fig. \ref{fig:1}(c). This is related to the phenomenon of \emph{Maker fringes} \cite{Maker}, which describe the z-dependence of the harmonic intensity for a monochromatic wave. In time domain, the solution of Eq.\,(\ref{eq:PulseProp1D}) predicts the generation of two well-separated pulses, where the leading pulse propagates at the speed of the IR pulse, whereas the trailing pulse propagates at the DUV group velocity [Fig. \ref{fig:1}(b)]. This pulse splitting, which is counter the intuition that a single broadened pulse might be generated, has been described before \cite{RN231}. For THG of pulse A, the leading pulse is labeled U and the trailing pulse is labeled V. At the end of the MgF$_2$ plate, U and V are well separated by $\sim$\,36\,fs with a pulse energy of $\sim$\,$6\ \mathrm{pJ}$ each. DUV light is also emitted collinearly to R and in the bisector between R and A for temporal overlap, but this DUV light is not detected.  

A scan of the pulse delay $\tau$ yields a data trace $\textbf{I}(\omega,\tau) = \left|U(\omega,\tau) + V(\omega,\tau) \right|^2$, see Fig. \ref{fig:1}(d). As a preparation for the pulse retrieval, the fringed spectra are inverse Fourier transformed, the side peak (alternating component) is cut-out and shifted to zero, and thereafter Fourier transformed. This yields $\textbf{I}_r(\omega,\tau) = U^*(\omega,\tau)V(\omega,\tau) \mathrm{e}^{-i \omega t_e}$, where $t_e$ represents the shift to zero which corresponds to the delay between U and V after the medium ($\sim$\,36\,fs for the present example of a 100-$\mu$m-thick MgF$_2$ plate). Pulse R interacts not at all with U or V for $\tau>50\ \mathrm{fs}$, see Fig. \ref{fig:1}(e). Since R propagates faster than V, R overtakes V somewhere inside the MgF$_2$ plate for $0<\tau<50\ \mathrm{fs}$ [Fig. \ref{fig:1}(b)]. For $20\ \mathrm{fs}<\tau<50\ \mathrm{fs}$, R interacts with V near the exit face of the MgF$_2$ plate. The overtaking is partial and hence the XPM of V depends on $\tau$. 

In the variant \emph{edge}, the pulse delay scan is restricted to a region where R is scanned solely over pulse V. The nonlinear medium must be thick enough that this restriction is possible. For the present example, the region $20\ \mathrm{fs} < \tau < 50\ \mathrm{fs}$ is selected, see Fig. \ref{fig:1}(e). In preparation for pulse retrieval, the trace
$\textbf{Y}_{\mathrm{e}}(\omega,\omega_{\tau})$
is calculated by differentiation of $\textbf{I}_r$ with respect to $\tau$ and a subsequent Fourier transform from $\tau$ to $\omega_{\tau}$. Since pulse U is not dependent on $\tau$ in the selected region of $\tau$, 
\begin{equation} 
\textbf{Y}_{\mathrm{e}}(\omega,\omega_{\tau}) = U^{\ast}(\omega)V'(\omega,\omega_{\tau})
\label{eq:Y1}
\end{equation} 
with $V' = \frac{dV}{d\tau}$. 
In order to develop a method for pulse retrieval, an analytic expression for $\textbf{Y}_{\mathrm{e}}(\omega,\omega_{\tau})$ is derived. The electric field of R is given by $R(\omega,\tau,z) = R(\omega) \mathrm{e}^{-i(\omega\tau+k(\omega)z)}$. This implies that R propagates linearly, in particular self-phase modulation (SPM) of R is neglected. The propagation of V is influenced by R through XPM, given for the case of instantaneous response by 
\begin{equation}
\begin{split}
P(\omega,\tau,z) =& 3\chi^{(3)}\iiint d\omega_1d\omega_2d\omega_3V_0(\omega_1)R(\omega_2)R(\omega_3) \\
  \times & \mathrm{e}^{-i(\omega_2+\omega_3)\tau}\mathrm{e}^{-i(k(\omega_1)+k(\omega_2)+k(\omega_3))z}\delta(\omega-\omega_{\sigma})
\end{split}
\label{eq:P3}
\end{equation}
with nonlinear susceptibility $\chi^{(3)}$, delta distribution $\delta(\omega)$ and $\omega_{\sigma} = \omega_1+\omega_2+\omega_3$. The pulse propagation [Eq.\,(\ref{eq:PulseProp1D})] for $E = V$ and $P$ given by Eq.\,(\ref{eq:P3}) is approximated with the assumption $n(\omega_2) = n(\omega_3) = n_{IR}$. After differentiation with respect to $\tau$ and Fourier transform to $\omega_{\tau}$, the result is given by 
\begin{equation} 
\begin{split}
V'(\omega,\omega_{\tau}) \propto & \phi(\omega) \omega \left(\mathrm{e}^{-i\left(\frac{n(\omega + \omega_{\tau})-n_{IR}}{c}\right) \omega_{\tau}z}-1\right)  \\
\times & V_0(\omega+\omega_{\tau})\mathcal{X}(-\omega_{\tau})
\end{split}
\label{eq:V4}
\end{equation}
with $\mathcal{X}(\omega_{\tau}) = \int d\omega_2R(\omega_2)R(\omega_{\tau}-\omega_2)$. In the following, the summand $-1$ in Eq. (\ref{eq:V4}) is neglected, because it corresponds to the interaction of R and V near $\tau = 0$ and hence must be excluded for the pulse delay range of \emph{edge}. The trace $\textbf{Y}_{\mathrm{e}}$, obtained by multiplication with $U^*(\omega)$, is given by
\begin{equation} 
\textbf{Y}_{\mathrm{e}}(\omega,\omega_{\tau}) \propto U^{\ast}(\omega) V(\omega + \omega_{\tau}) \mathcal{X}(-\omega_{\tau}).
\label{eq:Y6}
\end{equation}
The variant \emph{edge} is broadly applicable and not restricted to the present situation where the DUV pulses U and V are generated by the same VIS-IR pulse. U and V might be generated independently using any method and thereafter sent through the nonlinear medium for the XPM scan. The only restriction is that U and V must cover the same spectral region and that their time delay is sufficient to isolate the alternating part of the spectral fringes.  

In the variant \emph{center}, the pulse delay scan is restricted to a region where R is scanned over pulse U. For the present example, the region $-15\ \mathrm{fs} < \tau < 15\ \mathrm{fs}$ is selected, see Fig. \ref{fig:1}(e). The variant \emph{center} assumes that R and U travel with the same speed though the nonlinear medium. With regard to the scenario that U and V are DUV pulses and R is in the VIS-IR, this makes the applicability specific to the present DUV generation scheme. However, if all three pulses are frequency-degenerate, then R and U travel with the same speed though the nonlinear medium for any generation scheme and the variant \emph{center} is broadly applicable to any independently generated pulses U, V and R. It is assumed that XPM by R is only imposed on U and not on V, which is justified for the present case because U and R overlap temporally throughout the nonlinear medium, whereas R overtakes V within a propagation distance of $<10\ \mu\mathrm{m}$ for the given pulses. In preparation for pulse retrieval, the trace $\textbf{Y}_{\mathrm{c}}(\omega,\omega_{\tau})$ is calculated by subtraction of the non-$\tau$-dependent background from $\textbf{I}_r$, for example by subtracting $\textbf{I}_r(\omega,\tau = \infty)$ or $\textbf{I}_r(\omega,\tau = -\infty)$ from $\textbf{I}_r$ at each $\tau$, followed by Fourier transform from $\tau$ to $\omega_{\tau}$. Since pulse V is not dependent on $\tau$ in the selected region of $\tau$, the trace for retrieval is 
\begin{equation} 
\textbf{Y}_{\mathrm{c}}(\omega,\omega_{\tau}) = \left(U^{\ast}(\omega,\omega_{\tau}) - U^{\ast}(\omega,\tau = \infty)\right)V(\omega).
\label{eq:Y7}
\end{equation}
In contrast to the variant \emph{edge}, no differentiation is necessary, which is advantageous for noisy data. For data that is contaminated with noise, it may be beneficial to smooth the measured data over the $\tau$-axis before the numerical differentiation is carried out. In order to develop a method for analytic pulse retrieval, the pulse propagation [Eq.\,(\ref{eq:PulseProp1D})] for $E = U$ is solved with the approximation $n(\omega) = n(\omega_1) = n(\omega_2) = n(\omega_3) = n_{IR}$. After background subtraction and Fourier transform to $\omega_{\tau}$, the result is given by 
\begin{equation} 
\textbf{Y}_{\mathrm{c}}^*(\omega,\omega_{\tau}) \propto  \phi(\omega)  U(\omega + \omega_{\tau})V^*(\omega)\mathcal{X}(-\omega_{\tau}).
\label{eq:Y9}
\end{equation}

In both variants \emph{edge} and  \emph{center}, the data traces are of the advantageous kind that they factorize in three functions that only depend on $\omega$, $\omega_{\tau}$ and $\omega + \omega_{\tau}$ respectively. In Refs. \cite{Leithold1,Leithold2} an analytic as well as an iterative retrieval method was introduced for this kind of traces. Here, the analytic method is applied. For the numerical retrieval of $U$ and $V$, it is advantageous to perform the substitution $\omega_{\tau}$ $\rightarrow$ $\omega - \omega_{\tau}$ and to remove the fast phase oscillations of $\textbf{Y}_{\mathrm{e}}$ by multiplication with $\mathrm{e}^{-i \omega_{\tau} t_e}$. This results in
\begin{equation}
\bar{\textbf{Y}}_{\mathrm{e}}(\omega, \omega_{\tau}) \propto \mathrm{e}^{-i \omega_{\tau} t_e}U^{\ast}(\omega)V(-\omega_{\tau})\mathcal{X}(\omega_{\tau} + \omega)
\label{eq:Z6b}
\end{equation}
and
\begin{equation}
\bar{\textbf{Y}}^*_{\mathrm{c}}(\omega, \omega_{\tau}) \propto \phi(\omega) U(-\omega_{\tau})V^*(\omega)\mathcal{X}(\omega_{\tau} + \omega).
\label{eq:Z9b}
\end{equation}
As detailed in Refs. \cite{Leithold1,Leithold2}, preliminary solutions $\hat{U}(\omega)$ and $\hat{V}(\omega)$ are obtained by logarithmic differentiation with respect to $\omega$ respectively $\omega_{\tau}$ and subsequent exponential integration on the diagonal of the data trace. These are connected to the true solutions by 
$U(\omega) = \hat{U}(\omega)\mathrm{e}^{i \alpha + i T \omega + s \omega}$ and $V(\omega) = \hat{V}(\omega)\mathrm{e}^{i \beta + i T \omega + s \omega}$ with the real integration constants $\alpha, \beta, T$ and $s$. The absolute arrival time $T$ cannot be determined from the data traces, and also the determination of both carrier envelope phases $\alpha$ and $\beta$ individually is impossible. These parameters are sometimes referred to as \textit{trivial ambiguities}. The determination of $s$, which affects the spectra $\left|U(\omega)^2\right|$ and $\left|V(\omega)^2\right|$, is necessary for pulse retrieval. It is straightforward to determine $s$ if the spectrum of either U or V is known. Here, this is not the case and the spectrum has to be extracted from the double pulse spectrum $|U(\omega) + V(\omega)|^2$. For identical spectra of U and V, the spectrum can be deduced from the preliminary solutions using the identity $|U(\omega)|^2 = |V(\omega)|^2 = |\hat{U}(\omega)\hat{V}(-\omega)|$.

\begin{figure}[H] 
\centering
\includegraphics[width=\linewidth]{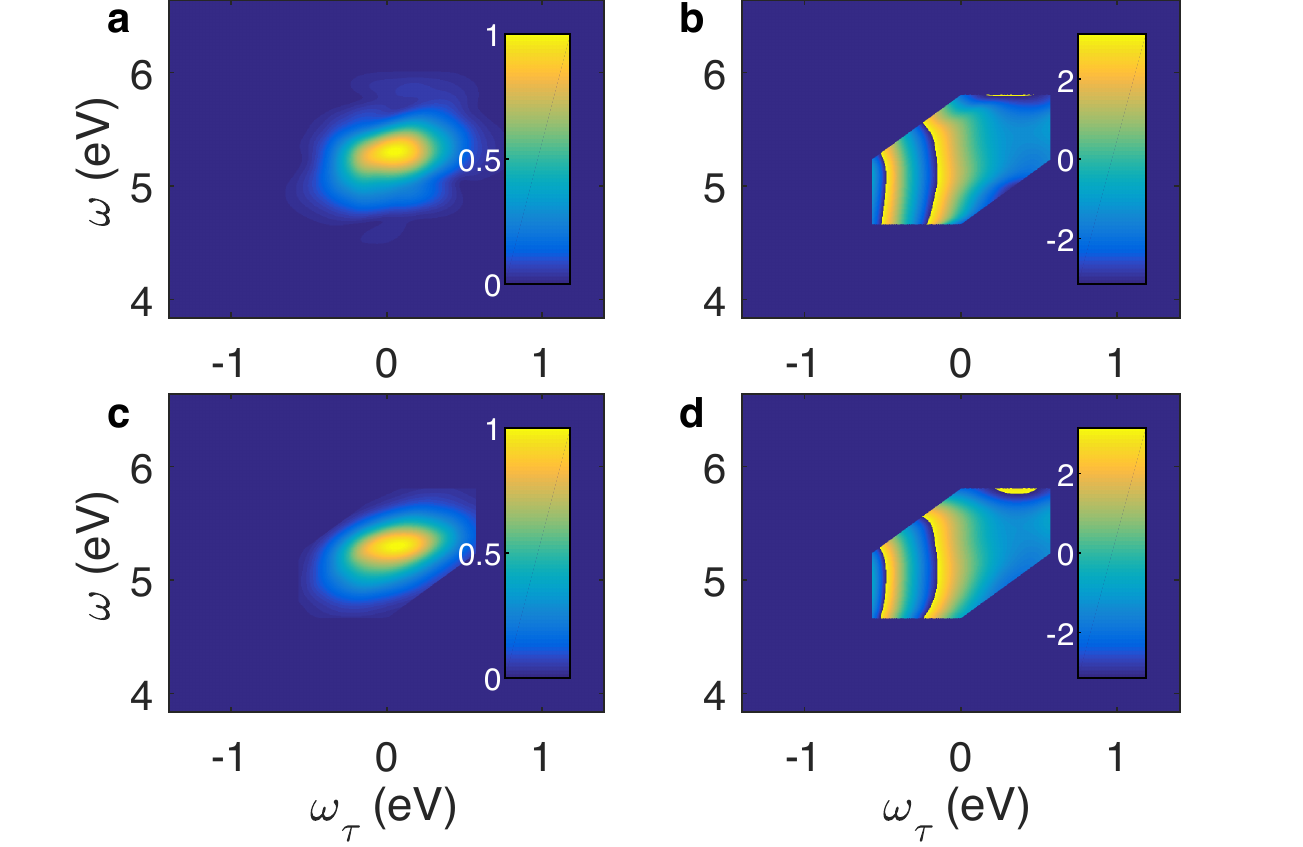}
\caption{Magnitude (a), (c) and phase (b), (d) of $\bar{\textbf{Y}}^*_{\mathrm{c}}(\omega, \omega_{\tau})$. The measured trace is depicted in (a) and (b), the retrieved trace in (c) and (d). The phase is only shown in the region where the magnitude exceeds 1\% of the maximum.}
\label{fig:2}
\end{figure}

\begin{figure}[H] 
\centering
\includegraphics[width=\linewidth]{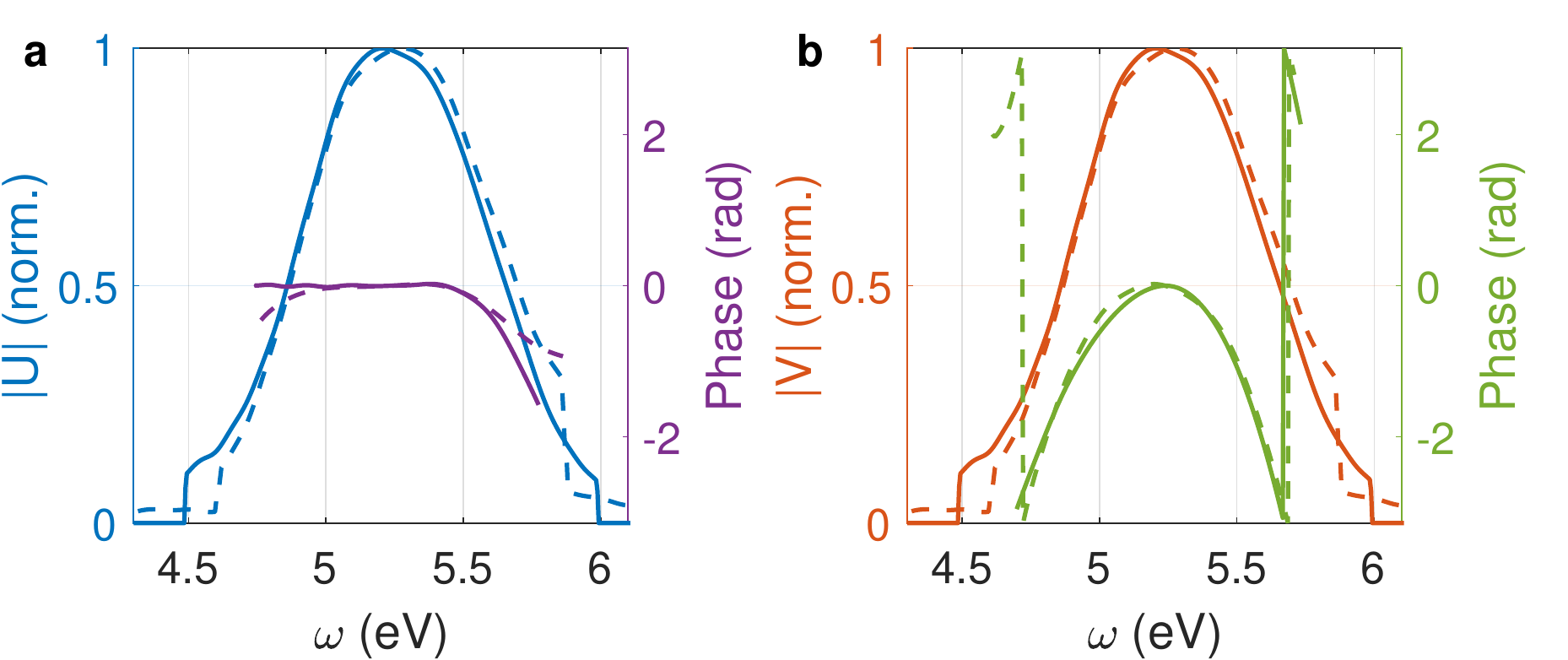}
\caption{The retrieved complex functions $U(\omega)$ (a) and $V(\omega)$ (b) using the methods \emph{edge} (solid) and \emph{center} (dashed). For better visibility, the linear phase is subtracted (corresponding to a shift to time zero) and the phase is displayed only for the region where the magnitude exceeds $20\%$ of the peak value.}
\label{fig:3}
\end{figure}

The pulse retrieval is applied to the experimental data trying both variants \emph{edge} and \emph{center}. In Fig. \ref{fig:2}, the magnitude and phase of the original data trace is shown for the variant \emph{center}. To suppress noise, all values below a threshold value of 1\% of the maximum value are set to zero. The retrieved functions $U(\omega)$ and $V(\omega)$ are shown in Fig. \ref{fig:3}. The results of \emph{edge} and \emph{center} are hardly distinguishable. The phase of U is almost flat as expected from the calculation. U is positively chirped by $12\ \mathrm{fs}^2$, which is due to the propagation inside the medium and also expected from the calculation. In the next step, the retrieved pulses are used to construct the retrieved trace using Eq.\,(\ref{eq:Z9b}). This provides a feedback to check the fidelity of the retrieval, similar as it is custom for FROG methods, with the advantage that both magnitude and phase values can be compared. The original and retrieved traces agree very well. The result for \emph{edge} (not shown) is almost identical. 

\begin{figure}[H] 
\centering
\includegraphics[width=\linewidth]{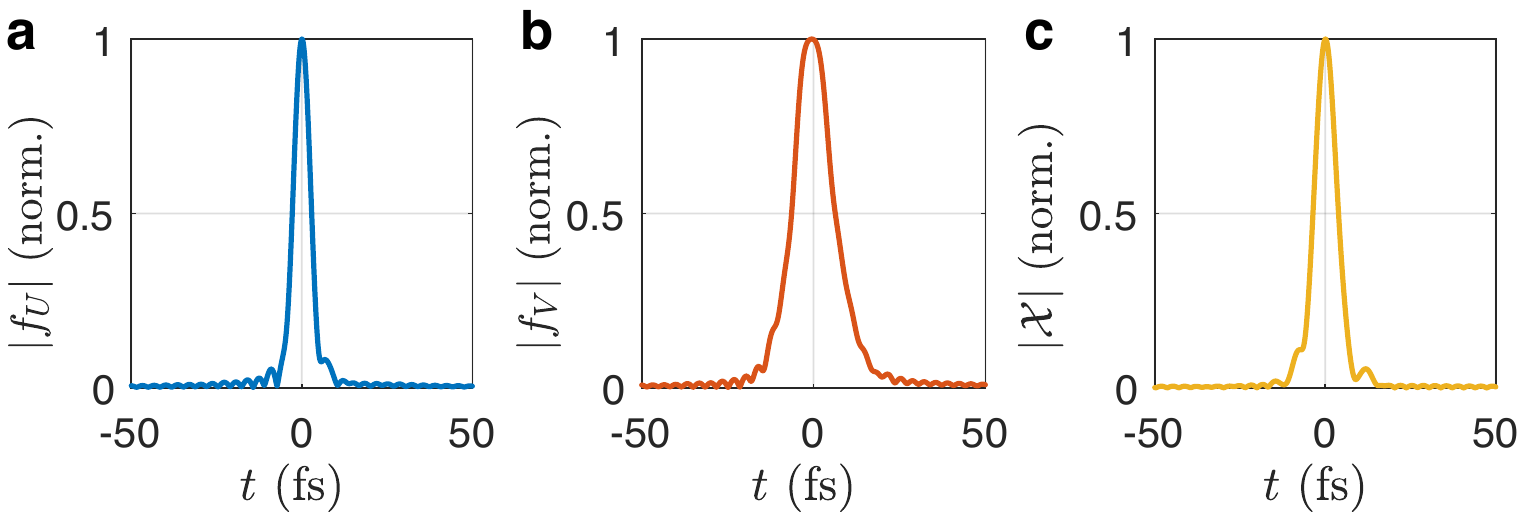}
\caption{Normalized magnitudes of the retrieved electric field envelopes $f_U(t)$ (a), $f_V(t)$ (\textbf{b}) and  $\mathcal{X}(t)$ (c). All pulses are shifted to time zero.}
\label{fig:4}
\end{figure}

For the time-domain representation of the pulses, the envelope functions $f_U(t) = \mathcal{F}^{-1} \{ U(\omega - \omega_0) \}$ and $f_V(t) = \mathcal{F}^{-1} \{ V(\omega - \omega_0) \}$ are calculated with the assumption of a center frequency $\omega_0$\,=\,5.2\,eV, see Fig.\,\ref{fig:4}. For U, the pulse duration $t^U_{FWHM} = 3.7\ \mathrm{fs}$ is obtained (FWHM of intensity envelope), which fits well to the estimation of THG in the limit of a thin sample ($t^U_{FWHM} =t^A_{FWHM}/\sqrt{3} = 3.5\ \mathrm{fs}$). The pulse duration of V is $t^V_{FWHM} = 9.1\ \mathrm{fs}$, which reflects the dispersive pulse broadening. The small oscillations underneath the pulse envelopes are numerical artifacts caused by the Fourier transform of the sharp edges where the trace is set to zero. They could be suppressed by using window functions. 

$\mathcal{X}(\omega)$ is also determined in the analytic pulse retrieval, which provides information about pulse R. The time-domain representation $\mathcal{X}(t) = \mathcal{F}^{-1} \{\mathcal{X}(\omega) \}$ gives an idea of the temporal shape of R, see Fig.\,\ref{fig:4}(c). $\mathcal{X}$ is the second-order process of R, however, it is not determined at the second harmonic but only near frequency zero (optical rectification).

\begin{figure}[H] 
\centering
\includegraphics[width=\linewidth]{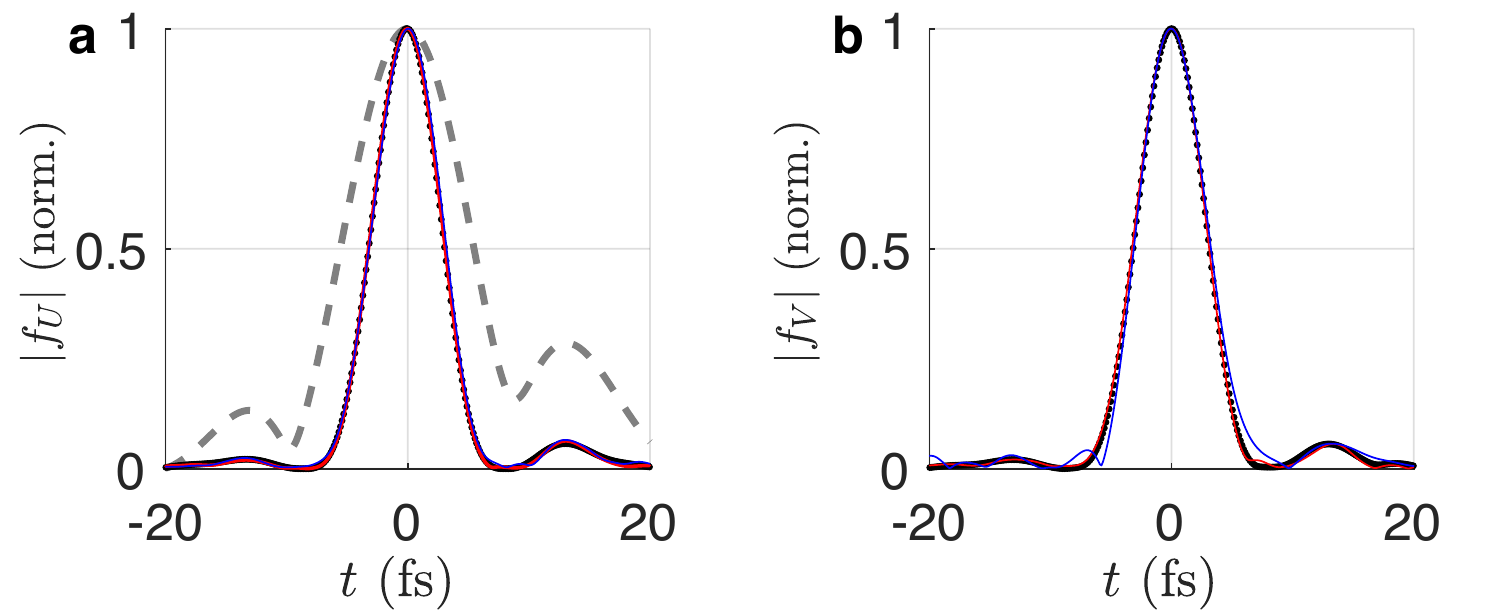}
\caption{Normalized magnitude of the envelope $f_U(t)$. The retrieved envelopes for the calculation with a step-like refractive index are depicted in (a), with numerical refractive index data of MgF$_2$ in (b). In (a) the variant \emph{edge} is used with $I_{R} = 0.1\,\mathrm{TW/cm}^2$ (red) and $I_{R} = 30\,\mathrm{TW/cm}^2$ (blue), in (b) the variants \emph{edge} (red) and \emph{center} (blue) are used with $I_{R} = 0.1\,\mathrm{TW/cm}^2$. The envelope of $f_A(t)$ and $f_R(t)$ is depicted for reference in (a) (gray-dashed), the envelopes of the original pulse U extracted from the calculations are depicted in both panels (black-dotted).}
\label{fig:5}
\end{figure}

In order to scrutinize aberrations of the retrieval, simulations are employed similar as described in Refs. \cite{Pati, Leithold1, Leithold2} with a randomly generated waveform for A and R (Fig.\,\ref{fig:5}). The unidirectional pulse propagation equation (UPPE) is integrated numerically using the split-step method. One transverse dimension (the $x$-dimension in Fig.\,\ref{fig:1}) is included to account for the noncollinear geometry. The electric field is treated as scalar field, because all pulses are polarized perpendicular to the plane of incidence. The nonlinear response is calculated assuming instantaneous response with $\chi^{(3)} = 0.82\ \mathrm{au}$, corresponding to a nonlinear refractive index n$_2$\,=\,0.057\,cm$^2$/PW \cite{DeSalvo}. The anticipated pulses have a non-trivial shape with pre- and post-pulses that extend over 40\,fs. In order to isolate the regions of the pulse delay scan for the variants \emph{edge} and \emph{center}, the thickness of MgF$_2$ is increased to  $150\,\mu$m. Subsequent to the sample propagation, linear propagation into the far field is performed and the DUV field in direction of A is extracted. First, a step-like refractive index is assumed with $n_{IR} = 1.39$ at frequencies of A and R and $n_{UV} = 1.50$ at frequencies of U and V. With this assumption, the solution of the $z$-integral in Eq.\,(\ref{eq:PulseProp1D}) is exact and the pulse retrieval using variant \emph{edge} for low intensity of R ($I_{R} = 0.1\,\mathrm{TW/cm}^2$) is nearly perfect [Fig.\,\ref{fig:5}(a)]. The FWHM of the intensity envelope is determined as 4.2\,fs. For higher intensity ($I_{R} = 30\,\mathrm{TW/cm}^2$), the neglect of SPM underlying Eq.\,(\ref{eq:P3}) causes slight aberrations and the retrieved pulse duration is 0.2\,fs longer compared to the original pulse. When numerical refractive index data of MgF$_2$ is used, the aberrations are also extremely low [Fig.\,\ref{fig:5}(b)]. For the variant \emph{center}, additional aberrations arise, because the assumption that XPM by R is only imposed on U and not on V is an approximation. Nevertheless the pulse retrieval is very good, and depending on the experimental conditions it is preferable to use variant \emph{center} in order to avoid the numerical differentiation. All aberrations found here cause that the retrieved pulses appear slighchartly longer than the original pulses. For a very precise pulse retrieval, simulations using experimentally retrieved pulses can be employed in order to correct for aberrations.

In conclusion, a method of pulse characterization using XPM scans was introduced. Two independent pulses U and V are retrieved by scanning a third unknown pulse R over either U or V. The variants \emph{edge} and \emph{center} adduce a broad applicability, including the cases that all three pulses are degenerate in frequency or not. Even energies of a few picojoules suffice for U and V, which is especially advantageous for DUV pulses. The pulse retrieval is analytic, ensuring rigour and speed, and at the same time a feedback from a retrieved trace is available to estimate its fidelity. 


\section*{Funding.}
Deutsche Forschungsgemeinschaft (DFG) (PF 887/1-1); Daimler und Benz Stiftung (32 07/15); Europ\"aische Fonds f\"ur regionale Entwicklung (EFRE) Th\"uringen (2016 FGI 0023); Carl-Zeiss-Stiftung.


\bibliography{lit_OL18}

\end{document}